\documentstyle [12pt]{article}
\input{tcilatex}
\begin{document}

\title{A History-Dependent Stochastic Predator-Prey Model : Chaos and its Elimination}
\author{Rouzbeh Gerami $^{1,2,}$ \footnote {rouzbeh@theory.ipm.ac.ir}
, Mohammad R. Ejtehadi $^{1,} \footnote{reza@theory.ipm.ac.ir}$}
\maketitle

{\it $^{1}$ Institute for Studies in Theoretical Physics and Mathematics,
 P.O.Box 19395-5531, Tehran, Iran}

{\it $^{2}$ Department of Physics, Sharif University of Technology,
P.O.Box 11365-9161, Tehran, Iran}

\begin{abstract}
A non-Markovian stochastic predator-prey model is introduced in which
the prey are immobile plants and predators are diffusing
herbivors. The model is studied by both mean-field approximation (MFA)and computer
simulations. The MFA results 
a series of bifurcations in the phase space of mean predator and
prey densities, leading to a chaotic phase. Because of emerging correlations
between the two species distributions, the interaction rate alters and if it 
is set the value which is obtained from the simulation, then the chaotic 
phase disappears.

\end{abstract}
PACS:

\section{Introduction}

The time evolution of systems of interacting species modeling natural
ecosystems has attracted wide attention since its first studies by Lotka 
\cite{Lotka} and Volterra \cite{Volterra}. Various models have been
introduced in order to consider different aspects of natural life, including motion,
birth and death processes, evolution and extinction \cite{Murray,
Vicsek,Penna}. Physical motivation for studying such models is that they exhibit 
interesting features such as chaos and critical phenomena.

A much studied category of such models is that of two interacting species,
the so called predator-prey systems \cite{Boccara,Satulovsky}. However, most of
the existing models neglect the effect of time delays on the dynamics of the models.
By time-delayed systems we mean such systems that their dynamics   
is not defined only by knowing their present state, but
some information about previous states is required. 
Time delays
are present in many different physical or biological systems, and are particularly
able to account for many features of ecological phenomena \cite{Murray,Faro,dd},
although they have not been studied extensively. 

In this paper we introduce a new model of the predator-prey
problem with history-dependent dynamics. In our model, herbivors 
and edible plants are the predators and the prey, respectively. The predators stray
randomly in a plant-full environment, eating them when they find any, but 
the eaten plant will regrow after a definite elapsed time. The predators
reproduce with a constant rate and die, if they have not eaten anything in a
specified length of time. In our model, time delays enter the temporal evolution
equations through the terms representing plants growth and predators death.
Somewhat similar models without such time delays were published before
\cite{Wolff,Mroz}.

The set of time-delayed equations leads to a rich collection of dynamical
behaviors including chaos. As we will show, however, the emerging space 
correlation of the densities can eliminate the chaotic behavior.

We have studied this model by discrete-time, lattice-based computer
simulations, as well as by a mean-field approximation solution. In what follows 
we describe our model and then present and discuss the results.

\section{The Model}

The ecosystem consists of a(n infinite) square lattice each site of which
if not empty, is occupied by either predators or a plant. The
predators move randomly to one of the nearest neighbors (two-dimensional
free random walk) and do not interact with each other, therefore multiple
occupancy of the site is allowed.
 If a predator
enters a site occupied by a plant, it will eat it. However after $c$ time 
steps another plant grows at that site.

To every predator an {\it energy} is assigned, indicating the
number of steps that it can go without eating anything. As a result, the energy is
lowered by one at every time step. Eating a plant raises the energy to the
maximum value $l$, so that a predator that has not eaten anything in $l$ steps will die. At
every time step each predator reproduces with probability $b$. The offspring is
positioned at the same site and half of the parent's energy is transferred
to it.

These rules are applied in the following order. The predators are first moved in a
random sequence. They eat every plant that they can, after which they
reproduce with some probability and finally plant growth occurs. In the case of
more than one predator entering a plant site, the early comer eats the plant.

The rules governing the motion of the predators are those that are characteristics of
branching diffusion processes, for which the space and time average
quantities as well as spatial correlations have been investigated \cite
{Althreya}. What we examine in the following sections is the time evolution
of the mean spatial densities of predators and plants.

\section{Fixed Points and Cluster Formation}

We first present the results obtained from simulating the
model. Simulations are made on a $M \times M$ square lattice with $M=100$ 
 and with periodic boundary conditions. As initial conditions, predators
and plants are distributed randomly and the value of $l$ is assigned to the energy 
of every predator.
The sites that are initially plant-free must be filled with plants in the first $c$
steps, so a random integer $\tau $, $0<\tau <c$, is assigned to every
such site, and a plant occupies that site at the $t=\tau$ 's time step .

Let $P({\bf x},t)$ and $N({\bf x},t)$ denote predator and plant respective
local densities and $p(t)$ and $n(t)$ be their respective spatial mean values,
i.e. $p(t) = \langle P({\bf x},t) \rangle$ and $n(t) = \langle N({\bf x}%
,t) \rangle$ (where $\langle . \rangle$ stands for spatial averaging). $P(%
{\bf x},t)$ is an integer number including $0$, while $N({\bf x},t)$ is either 
$0$ or $1$. $p(t)$ and $n(t)$ are assumed to be equal to the probability of
predators and plants occupying a lattice site (assuming that $p$ does not 
become larger than one).

As expected, time evolution of the two species can lead to a stationary
state (Fig. 1) in which both $n$ and $p$ fluctuate about their (time independent) mean values, and the fluctuations are predominantly out of phase.
Therefore by averaging $n(t)$ and $p(t)$ over many realizations of the system
 we find a fixed point in the $%
( \langle \langle p\rangle \rangle ,\langle \langle n\rangle \rangle) $ phase
space (Fig. 2) (where $\langle \langle .\rangle \rangle $ represents
the expectation value found by averaging over different realizations).

Trivially, $(n,p)=(1,0)$ is also a fixed point ({\it extinction} state). In
a wide range of parameters this is unstable, and there exists the just
described active oscillatory state with a $(\langle \langle p\rangle \rangle
,\langle \langle n\rangle \rangle)$ stable fixed point. But in a large
region in the parameter space of $l$, $b$ and $c$, the point $(1,0)$ is stable and there
is no non-extinction stationary state. This is the case for sufficiently
large $c$ (low growth rate for the plants), low $l$ (low energy content of a plant)
or low $b$ (low predator birth rate). Even an unstable fixed point, $%
(1,0)$ can be reached (in transient region) by specific initial conditions that
are large $p(0)$ or large $n(0)$. In the latter case, the initially high density
of plants increases $p$ and decreases $n$ very much and consequently all the predators
die of starvation. In the following we consider the non-trivial
(non-extinction) stationary state.

Although the predators (plants) have no interaction with each other, the spatial
distributions of $N({\bf x})$ and $P({\bf x})$ are not uniform in the stationary
state. This is due to the rules of the game that are random
motion of predators and the laws of birth and death \cite{Meyer}. As a
typical pattern, Fig. 3 shows the emergence of clusters of predators and
plants for $l=20$, $b=0.02$, $c=60$ and $t=500$, when the system is in its
stationary state. Formation of the clusters is characterized quantitatively by 
the predator or plant autocorrelation functions defined by
\begin{equation}
C_{n}({\bf d})=\frac{\langle N({\bf x}+{\bf d})N({\bf x})\rangle -n^{2}}{%
n^{2}}
\end{equation}
\begin{equation}
C_{p}({\bf d})=\frac{\langle P({\bf x}+{\bf d})P({\bf x})\rangle -p^{2}}{%
p^{2}}
\end{equation}
These clusters form separately, since if there is a plant at a site no
predators can be at the same site. This is shown by the predator-plant
correlation function:
\begin{equation}
C_{np}({\bf d})=\frac{\langle N({\bf x}+{\bf d})P({\bf x})\rangle -np}{np}
\end{equation}
Fig. 4 shows $\langle \langle C_{n}({\bf d})\rangle \rangle $, $\langle
\langle C_{p}({\bf d})\rangle \rangle $ and $\langle \langle C_{np}({\bf d}%
)\rangle \rangle $ as functions of $d$ along the lattice axis , 
for the same parameter set as in Fig.3. They all
vanish as $d$ increases, but while $C_{n}$ and $C_{p}$ are positive
functions for small $d$, representing formation of the clusters, $C_{np}$ is 
negative since the
probability that a plant occupies a site decreases if there is a predator
in the neighborhood. Diffusion of the predators increases the fluctuations in $C_{p}$
and $C_{np}$. An exponential function best fits to $C_{n}$ with correlation
length increasing with $c$.

\section{Mean-Field Approximation}

We consider the correlation of the predators and plants within a mean-field
approximation. If the probabilities that a site is occupied by a
predator or a plant were independent, the density of the eaten plants at
every time step would be given by 
\begin{equation}
\Delta _{-}n(t)=n(t)p(t)
\end{equation}
i.e. $\Delta _{-}n(t)$ is the        
probability of that a site is simultaneously occupied by both a
predator and a plant. To take into account the just described correlations,
we modify this expression, by writing as 
\begin{equation}
\Delta _{-}n(t)=rn(t)p(t)
\end{equation}
where $0<r<1$ and also can be thought of as a rate. Stronger correlations 
imply larger clusters which lowers
the value of $r$. Introduction of $r<1$ rate, can also be justified
in this way: since predators move randomly, a predator lowers its
food-eating chance by repeatedly coming back to the sites which had previously 
been occupied by itself and it had eaten the plants in them. We
show that $r$ is an important
parameter that controls the ability of the system to transit into a chaotic phase.

To calculate $r(t)$ by simulation, we enumerate the total number of the eaten
plants at time $t$ and divide it by $M{^{2}}n(t)p(t)$. Figure 5 represents
as a function of time, the value of $\langle \langle r(t)\rangle \rangle $
for $l=20$, $b=0.02$ and $c=60$ which indicates that it becomes essentially a
constant at about $\langle \langle r(t)\rangle \rangle \simeq 0.54$ in the
stationary state. In fact $\langle \langle r(t)\rangle \rangle $ varies slightly
as $a$, $b$ and $c$ change.
The value of $r$ can also be read from the correlation function (Fig 4c).
Since the probability that a plant is eaten is $\frac{1}{4}$ of probability
that a predator and a plant are nearest neighbors, and this probability is equal to the
probability of finding a predator and a plant within a unit distance {\bf i}, 
we have 
\begin{equation}
\Delta _{-}n=rnp=\langle n({\bf x}+{\bf i})n({\bf x})\rangle
\end{equation}
then 
\begin{equation}
C_{np}({\bf i})=r-1.
\end{equation}

>From Fig. 4(c) we find that $C_{np}({\bf i}) \simeq 0.46$ and $r=0.54$ in
complete agreement with the independently calculated value of $r$ (Fig. 5).

The time evolution equations will then be 
\begin{eqnarray}
n(t+1)&=&n(t)+r[n(t-c+1)p(t-c+1)-n(t)p(t)] \\
p(t+1)&=&p(t)\left\{ (1+b)-\prod_{t^{\prime }=0}^{l-1}[1-rn(t-t^{\prime
})]\right\}.
\end{eqnarray}

The second term on the right-hand side of Eq. (8) is $\Delta _{-}n(t)$ and
the first is $\Delta _{+}n(t)=\Delta _{-}n(t-c+1)$, the density of the plants eaten
at time $(t-c+1)$ which will grow again after $c$ steps at time $t+1$. The second
term on the right-hand side of Eq.(9) is the probability of a predator not
eating anything in each of the past $l$ steps. At any time, a predator does not
eat a plant with a probability

\begin{equation}
\frac{ p(t)-rn(t)p(t) }{ p(t) } = 1-rn(t) 
\end{equation}
i.e. the ratio of density of those predators who do not eat to the total predators density.

\section{Solution of the Mean-Field Equations}

In order to find the possible solutions of this set of equations, we numerically
compute $p(t+1)$ and $n(t+1)$, knowing the values of $p$ and $n$ at the earlier
times. By repeatedly doing this, we can find all the possible trajectories in the $(p,n)$
phase space. However, because of the existence of time delays, it is not
sufficient to know only $p$ and $n$ at time $0$ in order to initiate these equations.
Therefore, to overcome this difficulty we rewrite these equations as

\begin{eqnarray}
n(t+1)&=&n(t) + \frac{1-n(0)}{c} - rn(t)p(t) \hspace{2cm} {\rm for} \ \   t < c\\
p(t+1)&=&p(t) (1+b)      \hspace{5cm} {\rm for} \ \  t<l
\end {eqnarray}
We drop the $rn(t-c+1)p(t-c+1)$ term from Eq. (8) for $t<c$, and add the term 
\begin{equation}
\frac{1-n(0)}{c}
\end{equation}
 to take into account growing of the plants in the initially
plant-free sites. Also, we eliminate the second term in Eq. (9) for $t<l$
because of the initially full energy of all the predators.

As in the case of the simulations, $(1,0)$ is a trivial fixed point, which is 
unstable only for sufficiently low $l$, low $b$ or high $c$ and long-time
behavior of the solutions does not
depend on the values of $n(0)$ and $p(0)$. Here, we do not consider that range of
the parameters for which extinction occurs. In order to find a fixed point
(Fig. 6) we assume that $n$ and $p$ are constant for a long time, that is $%
n(t^{\prime })=n^{*}$ for $t-l\leq t^{\prime }\leq t$ and $%
p(t-c+1)=p(t)=p^{*}$. This leads to

\begin{equation}
n^{*} = \frac{1}{r} [1-(1+b)^{1/l}]
\end{equation}
(which is independent of $c$ ) but no explicit expression for $p^{*}$. However,
interestingly, as we have checked numerically, there also exists a unique $%
p^{*}$ and the fixed point (specially its $p$-coordinate) is uniquely 
determined by the parameters and independent of initial 
conditions.  

Although $r$ is obtained definitely from simulation, we assume it to be
variable. It is easily seen that the dynamics of the equations depends
critically on the value of $r$. We temporarily assume $r=1$ which means
neglecting the correlations and clustering.

For every $l$ and $b$ the fixed point is stable for low $c$. As $c$ is
increased, the fixed point eventually loses its stability through a Hopf bifurcation and
turns to a limit cycle (Fig. 7). As $c$ is increased further, more bifurcations
occur which lead to the chaotic phase (Fig. 8). Figure 9 represents the
bifurcation diagram, which is the Poincare maps for constant $l$ and $b$ and
varying $c$, obtained from the intersection of trajectories in the phase space 
with the vertical line $n=n^{*}$. Fig. 10
is the same graph for the same set of parameters except that $r=0.48$ (see
below), in which the chaotic regime has been eliminated.

We found that the time-evolution equations with a realistic value
of $r(l,b,c)$ which is obtained from the simulation, do not exhibit a chaotic
behavior. This is similar to what occurs in the simulation which always there 
exists a stationary
state. However, setting $r=1$ artificially, can produce a chaotic
behavior. This has an interesting interpretation: formation of the clusters and
emergence of correlations removes the chaotic regime.

Finally, Fig. 11 offers a comparison between the mean-field results and those of
the computer
simulations. Here, the fixed points in $(\langle\langle p \rangle\rangle,
\langle\langle n \rangle\rangle)$ phase space derived from simulation and
the time-average of $(p,n)$ 	
obtained from the mean-field equations are shown with $%
l=20$, $b=0.02$, and varying $c$ from 65 to 150. $r$ is chosen so that the two
curves best coincide and that occurs if $r \simeq 0.48$.

\section{Conclusion}

We have introduced and studied a model for the predator-prey problem with time delay in
which the prey are edible plants and the predators are herbivores. The model
is defined algorithmically through a series of rules that are
i) random-walk motion of the predators; ii) growth of the eaten plants after a time delay,
and iii) death of those predators not having eaten anything in a specified length
of time. Both rules ii and iii generate history dependence in the mean-field
equations. Simulation of the model on a lattice yields
stationary states with fixed points in the phase space of $(\langle \langle
p\rangle \rangle ,\langle \langle n\rangle \rangle)$ as well as a trivial $%
(1,0)$ fixed point.

In such stationary states, the predators and the prey are distributed in
separately-formed clusters and hence producing non-zero autocorrelation as well
as correlation functions. Such correlations are taken into account in the
mean-field equations by introduction of a rate $r<1$ in
the expression of the eaten plants density. These equations have chaotic
solutions for $r$ nearly 1, but there is no
chaos if $r$ is lowered to its true value obtained from the simulation.

The authors offer their special thanks to N.Hamedani, V.Shahrezaei, H.
Seyed-Allaei and S.E. Faez for invaluable discussions and to M. Sahimi, A.Erzan and 
A.Aghamohammadi for carefully reading the manuscript and useful comments.

\newpage

\begin{center}
{\Large Figure Captions}
\end{center}

{\bf Figure 1.}

Predator and plant mean densities with respect to time for $l=20$, $b=0.02$, $%
c=80$, $p(0)=0.01$ and $n(0)=0.75$. (a) $p(t)$, (b) $n(t)$ \vspace{1cm}

{\bf Figure 2.}

Fixed point in the phase space of expectation values of predator and plant
mean densities $(\langle \langle p\rangle \rangle ,\langle \langle n\rangle
\rangle $), for the same parameter set as in Fig. 1.

\vspace{1cm}

{\bf Figure 3.}

Distribution of predators and plants in a $100 \times 100$ lattice for $%
l=20$, $b=0.02$, $c=60$ and $t=500$. Predators are represented by black dots and
plants by grey.

\vspace{1cm}

{\bf Figure 4.}

expectation values of (a) autocorrelation function of predators (b)
autocorrelation function of plants (c) predator-plant correlation function,
as a function of $d$ along the lattice axis for $l=20$, $b=0.02$, $c=150$ and $t=500$. An exponential function best fits to $C_{n}$ with correlation length
increasing with $c$.

\vspace{1cm}

{\bf Figure 5.}

Numerically calculated expectation value of $r(t)$ as a function of $t$ for $%
l=20$, $b=0.02$ and $c=60$, with $0.54$ mean value for $t>400$.

\vspace{1cm}

{\bf Figure 6.}

A fixed point in $(p,n)$ phase space, derived from mean-field equations, for 
$l=20$, $b=0.02$ and $c=38$.

\vspace{1cm}

{\bf Figure 7.}

A limit cycle in $(p,n)$ phase space, derived from mean-field equations, for 
$l=20$, $b=0.02$ and $c=67$.

\vspace{1cm}

{\bf Figure 8.}

Long-time behavior of other possible solutions of mean-field equations, as $%
c $ is increased with other parameters constant (transients have been
omitted): (a) two cycles for $c=75$ (b) four cycles for $c=82$ (c) chaos for 
$c=84$ , in this case the total area is filled in the long times.

\vspace{1cm}

{\bf Figure 9.}

Bifurcation diagram: intersection points of trajectories in the phase space 
with vertical line $n=n^{*}$, for
the same $l$ and $b$ as Fig.7 and $r=1$.

\vspace{1cm}

{\bf Figure 10.}

The same as Fig. 9 but with $r=0.48$

\vspace{1cm}

{\bf Figure 11.}

Comparison of simulation and mean-field results: black squares are
numerically calculated fixed points in $(\langle\langle p
\rangle\rangle,\langle\langle n \rangle\rangle)$ phase space. White squares
are time average of $p$ vs. time average of $n$ derived from mean-field
equations. $r$ is chosen to be $0.48$ which best coincides two curves.

\vspace{1cm}


\begin{thebibliography}{99}
\bibitem{Lotka}  A.J.Lotka, Proc Natl. Acad. Sci. U.S.A. {\bf 6}, 410 (1920)

\bibitem{Volterra}  V.Volterra, Mem. Accad. Nazionale Lincei 2, {\bf 6} , 31
(1926)

\bibitem{Murray}  J.D.Murray, {\it Mathematical Biology} (Springer Verlag,
N.Y., 1993)

\bibitem{Vicsek}  Z.Csahok , T.Vicsek, Phys.Rev.E. {\bf 52}, 5297 (1995)

\bibitem{Penna}  T.J.P.Penna, J.Stat.Phys. {\bf 78}, 1629 (1995)

\bibitem{Boccara}  N.Boccara, O.Roblin, M.Roger , Phys.Rev.E. {\bf 50} ,
4531 (1994)

\bibitem{Satulovsky}  J.Satulovsky, T.Tome , Phys.Rev.E. {\bf 49} , 5073
(1994)

\bibitem{Faro}  J.Faro, S.Velasco, Physica D, {\bf 110} (1997), 313

\bibitem{dd}  V.Mendez, J.Camacho, Phys.Rev.E. {\bf 55} (1997), 6476

\bibitem{Wolff} W.F.Wolff, page 285 in {\it Ecodynamics - Contributions to 
Theoretical Ecology}, edited by W.Wolff, C.J.Soeder, F.R.Drepper, (Springer,
Berlin-Heidelberg, 1988).

\bibitem{Mroz} I.Mroz and A.Pekalski, Eur.Phys.J. B {\bf 10}, 181 (1999).

\bibitem{Althreya}  K.B.Althreya and P.E.Ney, {\it Branching Processes}
(Springer, Berlin, 1972).

\bibitem{Meyer}  M.Meyer, S.Havlin and A.Bunde, Phys.Rev.E. {\bf 54}, 5567
(1996).
\end{thebibliography}
\end{document}